\documentclass{blois}

\usepackage{amssymb}
\usepackage{amsmath}

\def\be{\begin{equation}}
\def\ee{\end{equation}}
\def\bea{\begin{eqnarray}}
\def\eea{\end{eqnarray}}

\newcommand{\Photo}{\vspace{5mm} \includegraphics[height=35mm]{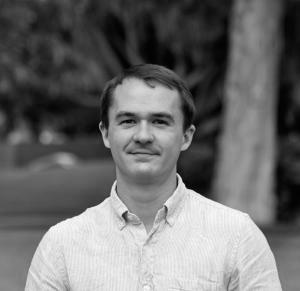}}

\begin{document}
\vspace*{4cm}
\title{IMPLICATIONS OF SUSY SEARCHES FOR PHYSICS BEYOND THE STANDARD MODEL}

\author{Nathaniel Craig}

\address{Department of Physics, University of California, Santa Barbara, CA 93106}

\maketitle\abstracts{
I discuss some essential features of the electroweak hierarchy problem and the ensuing motivation for weak-scale supersymmetry. Taking the hierarchy problem seriously, null results in searches for SUSY at the LHC favor specific regions of SUSY parameter space. More broadly, they suggest investigating a variety of alternative approaches to the hierarchy problem with diverse experimental signatures.}

\section{The Higgs and its Hierarchy Problem}

The discovery of the Higgs boson was the great triumph of the first run of the LHC. The great challenge for the second run will be to understand why it is so light. Observationally, the Higgs doublet mass parameter is $m^2 \sim (89 \, {\rm GeV})^2$; in the context of the Standard Model, this is simply an experimental scale. But it's a very curious scale at that. We know that the Standard Model is itself not a complete description of nature; at the very least there are also gravitational interactions, which as a quantum field theory are intrinsically non-renormalizable with a corresponding scale $M_{P} \sim 10^{19}$ GeV. The gravitational field theory contains an infinite number of irrelevant operators equally important at the scale $M_P$, and presumably is completed by a more fundamental description above $M_P$.

The hierarchy between these scales is worrisome; we have two dimensionful scales that differ by more than sixteen orders of magnitude. However, we already see a hierarchy of nearly six orders of magnitude in the known fermion masses (between the electron at 0.5 MeV and the top quark at 173 GeV), or something in the ballpark of twelve orders of magnitude between neutrino masses and the top quark. One wonders if one hierarchy is really any more troubling than the others. 

\begin{figure}[t] 
   \centering
   \includegraphics[width=3in]{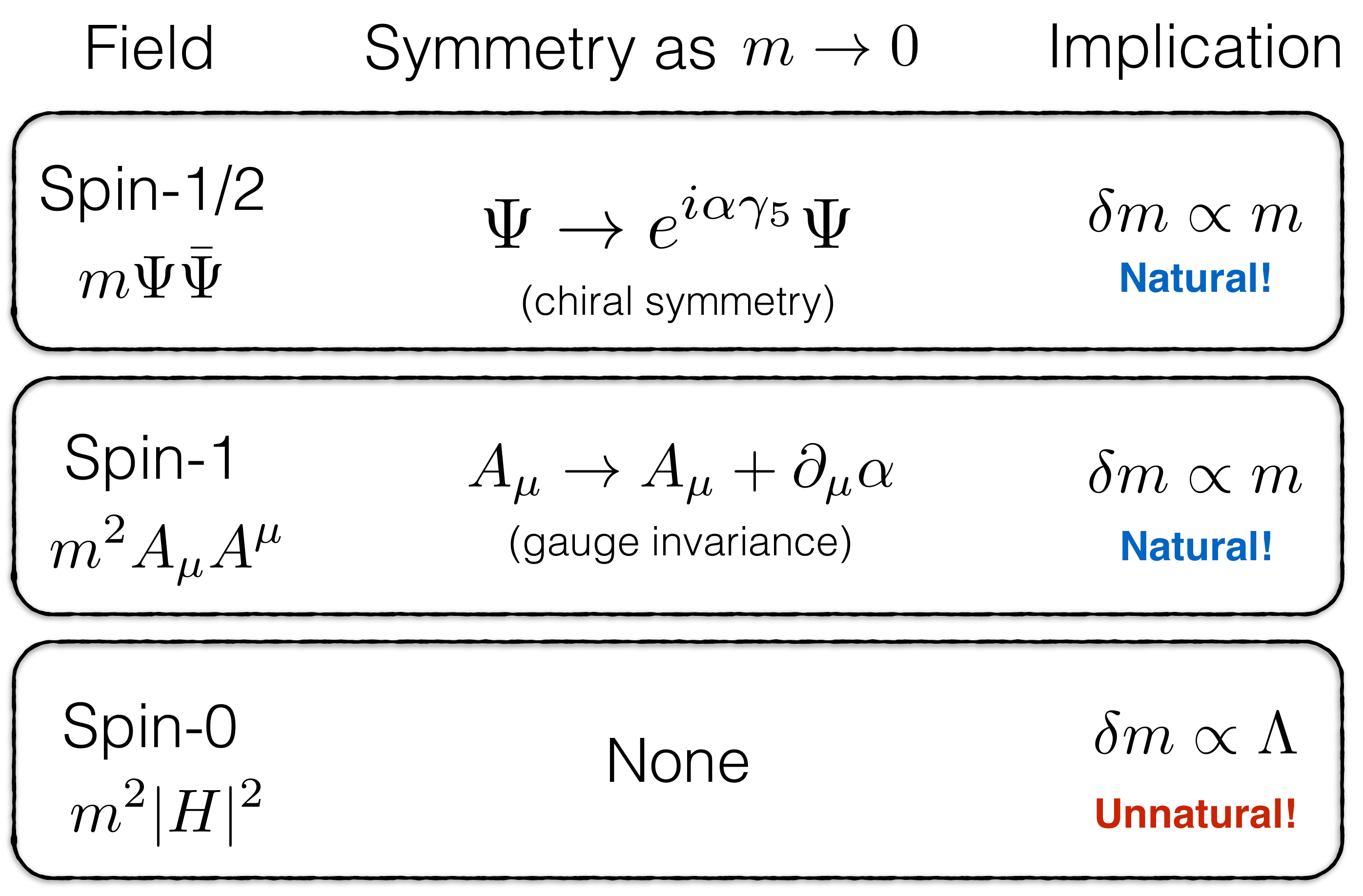}
   \caption{Custodial symmetries (or the lack thereof) for the mass terms of fermions, vector bosons, and scalars. The lack of an enhanced symmetry as $m^2 \to 0$ is the heart of the hierarchy problem for the Higgs.}
   \label{fig:spins}
\end{figure}

The key difference arises at the quantum level. In the case of fermion masses, the hierarchy is merely a just-so story. Consider a Dirac fermion $\Psi$ with a mass term of the form
\begin{equation}
m \bar \Psi \Psi \, .
\end{equation}
This mass term is invariant under global abelian rotations of the form $\Psi \to e^{i \alpha} \Psi$, but in the limit $m \to 0$ there is an additional symmetry, namely axial transformations of the form $\Psi \to e^{i \alpha \gamma_5} \Psi$. Quantum corrections respect the symmetries of the quantum action, so when $m = 0$ this implies that quantum corrections will not generate a mass term. Moreover, when the chiral symmetry is broken by $m \neq 0$, quantum corrections will be proportional to the symmetry-breaking term. Thus a large hierarchy between fermion masses is a curiosity, but not a deeply troubling one. If the fundamental theory of the universe generates fermions with very different masses, quantum corrections need not disturb the hierarchy. This is a beautiful property not only of spin-1/2 particles, but of spin-1 particles as well. In the case of vector bosons, in the limit where the mass term
\begin{equation}
m^2 A_\mu A^\mu
\end{equation}
goes to zero, there is an enhanced symmetry -- gauge invariance under $A_\mu \to A_\mu + \partial_\mu \alpha$. This likewise guarantees that radiative corrections to gauge boson masses are proportional to the mass itself. (In truth the situation is a bit more subtle for gauge bosons, but this broad-brush treatment is sufficient for our current purposes.) These are {\it custodial symmetries} for the mass parameters of spin-1/2 and spin-1 fields, explaining how such fields can be insensitive to hierarchies of scale.

The same does not in general hold for the mass terms for scalar fields. The mass term
\begin{equation}
m^2 H^\dag H
\end{equation}
is in general a complete invariant under any gauge or global symmetry acting on $H$. In particular, no symmetry is enhanced when the mass is zero. Thus we are without any argument to justify the stability of the Higgs mass parameter against radiative corrections. Indeed, we find in any theory with multiple mass scales that the Higgs accumulates radiative corrections from every scale with which it interacts, proportional to those scales. Unlike the case of spin-1/2 or spin-1, we do not have $\delta m^2 \propto m^2$, but rather $\delta m^2 \propto \Lambda^2$, where $\Lambda$ stands for all other scales probed by the Higgs. While we often speak of this sensitivity in terms of quadratic divergences in an effective field theory with a hard momentum cutoff $\Lambda$, in truth it is independent of regularization and renormalization scheme; in mass-independent schemes the effects are simply threshold corrections. 

We have many reasons to expect a variety of physics above the weak scale, and at the very least must contend with the physics of quantum gravity at $M_P$. While it is tempting to try avoiding problems by positing the absence of scales above the weak scale, this asks an absurd amount from the (unknown) physics of quantum gravity. Moreover, if there are no physical scales above the weak scale and quantum gravity does not provide a cutoff at $M_P$, the Standard Model itself generates a scale when the $U(1)_Y$ gauge coupling grows strong in the far UV.

Rather than relying upon unspecified ultraviolet miracles, a more conservative strategy is to resolve the hierarchy problem by extending the Standard Model. The essential idea is to enlarge Standard Model so that Higgs also enjoys a custodial symmetry, much as the fermions and gauge bosons. Note that this was not historically the only way to solve hierarchy problem. For example, we could have imagined lowering the scale of gravitational physics, as in theories with large extra dimensions. Alternately, we could have imagined breaking electroweak symmetry with some sort of strong condensate, as in technicolor. But now we have seen a light, apparently elementary Higgs scalar responsible for electroweak symmetry breaking, with an apparent separation of scales between the Higgs and any new physics. Therefore our focus narrows to UV physics that solves hierarchy problem for a light, approximately elementary scalar; this leads us to focus on custodial symmetries. 

What are the possible symmetries can we use? The Coleman-Mandula theorem \cite{Coleman:1967ad} (and its generalization by Haag, Sohnius, and Lopuszanski \cite{Haag:1974qh}) constrains the possible options to include internal symmetries with spinorial charges (supersymmetry); internal symmetries with scalar charges (global or gauge symmetry); and conformal symmetry.
\section{The case for SUSY}
Of these options, supersymmetry is often considered the most attractive, and with good reason. SUSY accomplishes everything we want from a custodial symmetry for the Higgs mass. Although there are many ways to think of the sense in which supersymmetry solves the hierarchy problem, a simple one is to observe that supersymmetry places the Higgs scalar into a supermultiplet with a Higgs fermion, the Higgsino. The mass of the Higgsino is itself protected by chiral symmetry; since $\delta \mu \sim \mu$ for the Higgsino, same holds from Higgs. Of course, supersymmetry is not an exact symmetry of the weak scale and must be broken -- but if it is broken by soft terms, then radiative corrections due to supersymmetry breaking must be proportional to these terms and UV sensitivity is avoided. Now the Higgs mass can be calculated in terms of the soft masses of states with which it interacts.

Although most strongly motivated as a solution to the hierarchy problem, supersymmetric extensions of the Standard Model exhibit a variety of other virtues. They often furnish a viable dark matter candidate, though the most conventional supersymmetric candidates are under increasing strain from direct and indirect detection. The additional SM-charged matter predicted by supersymmetric extensions typically improves the prediction for precision gauge coupling unification. Minimal extensions predict an elementary Higgs boson below $\sim 135$ GeV or so, in excellent agreement with observation. Finally, although SUSY theories extend the particle content of the Standard Model, these additional degrees of freedom have a well-behaved limit in which corrections to Higgs couplings and precision electroweak observables decouple. In this respect, SUSY theories are in good agreement with the apparently Standard Model-like character of the weak scale.

Having rendered the Higgs mass finite and calculable with supersymmetry as our custodial symmetry, we can develop expectations for the mass spectrum of superpartners based on their contributions to the Higgs mass parameter. The dominant contributions to the Higgs mass parameter come from three places. The first is a tree-level contribution proportional to the Higgsino mass, as required by supersymmetry. Avoiding excessive contributions to the Higgs mass requires the Higgsinos to be light, ideally in the ballpark of 200 GeV or lighter. The second appreciable correction arises from one-loop threshold corrections. By far the largest correction of this form is due to the stops, since the top chiral superfields couple most strongly to the Higgs, with a correction of order 
\begin{equation}
\delta m_{H}^2 = - \frac{3 y_t^2}{4 \pi^2} m_{\tilde t}^2 \ln \left(\Lambda / m_{\tilde t} \right)
\end{equation}
Avoiding excessive contributions to the Higgs mass parameter requires stops $\sim 400$ GeV with a cutoff $\Lambda \sim 10$ TeV. Other states also contribute at one loop, but their threshold corrections are proportional to smaller couplings and hence the states can be heavier without implying a tuning. 
The third appreciable contribution is from two-loop threshold corrections, as the Higgs interacts with gluinos at two loops with sizable coefficients that partially compensate for the additional loop factor. One can think of these corrections as coming from the gluino contribution to the stop mass, of order
\begin{equation}
\delta m_{\tilde t}^2 = \frac{2 g_s^2}{3 \pi^2} m_{\tilde g}^2 \ln \left(\Lambda / m_{\tilde g} \right)
\end{equation}
This ties $m_{\tilde g} \lesssim 2 m_{\tilde t}$, and implies that gluinos cannot be far from the mass scale of the stops.

Although supersymmetric extensions of the Standard Model involve a zoo of new particles, these naturalness considerations provide a useful way of organizing our expectations for the mass spectrum. It provides a sort of minimal set of expectations for where states should lie if supersymmetry solves the weak hierarchy problem without undue tuning. Of course, we should keep in mind that the details of UV physics may adjust these expectations, and also that superpartners irrelevant for naturalness may still be light. This arrangement of expectations according to the size of threshold corrections has a long history,\cite{Dimopoulos:1995mi}$^,$ \!\cite{Cohen:1996vb} but more recently has come to characterize the paradigm of ``Natural SUSY.'' \cite{Papucci:2011wy}

\section{SUSY Searches at the LHC and Their Implications}

\begin{figure}[t] 
   \centering
   \includegraphics[width=2.6in]{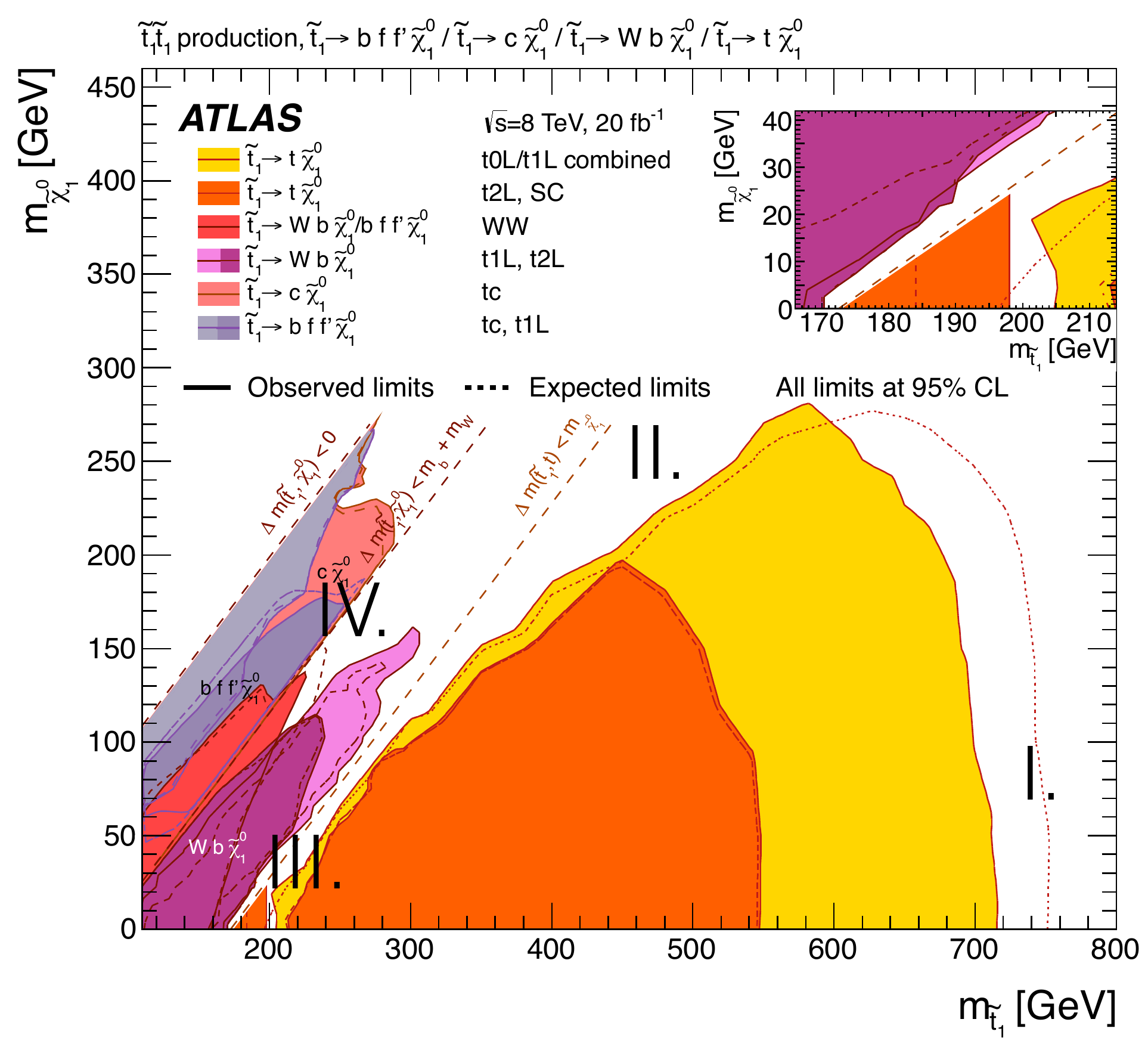} 
    \includegraphics[width=3.5in]{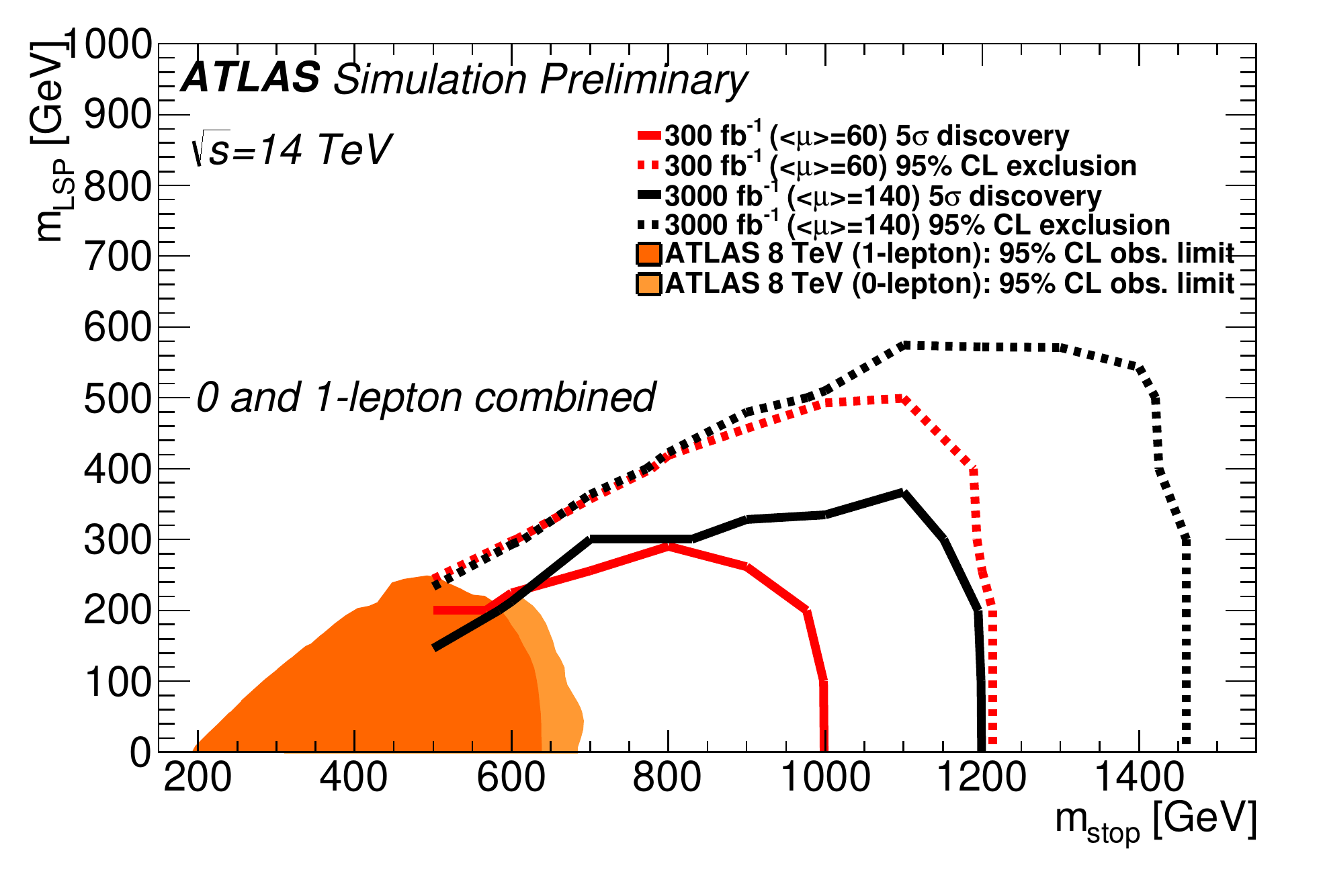} 
   \caption{Left: Summary of Run 1 stop searches at ATLAS; CMS limits are comparable. Regions labeled I.-IV. denote interesting kinematic regimes discussed in the text. Right: ATLAS simulated reach in stop searches at $\sqrt{s} = 14$ TeV. }
   \label{fig:stops}
\end{figure}

Of the superpartners that must be light according to naturalness considerations, the best limits may be set on states carrying QCD quantum numbers. Searches for supersymmetry at the LHC have largely proceeded along these lines, with considerable emphasis on searches for the production and decay of stops and gluinos. The current state of stop searches is summarized in Fig.~\ref{fig:stops} (here I have shown the ATLAS summary plot;\cite{Aad:2015pfx} the CMS reach is comparable). These limits assume that the stop decays into various Standard Model states plus an invisible LSP $\tilde \chi_1^0$. In generic regions of parameter space -- labeled region I in Fig.~\ref{fig:stops} -- the limits now reach $m_{\tilde t} \sim 700$ GeV. If stops do indeed lie above this range, it's cause for some concern; absent any special structure to the UV theory, this implies that the weak scale is tuned to the 10\% level. We have not shown the comparable limits on gluinos; in generic regions of parameter space the gluino limits reach $m_{\tilde g} \sim 1.4$ TeV. Although the gluino bounds are stronger, they are compatible with stop limits in the sense that they still satisfy the naturalness relations discussed above. Looking forward to the ATLAS simulated reach at $\sqrt{s} = 14$ TeV,\cite{ATL-PHYS-PUB-2013-011}  we see that the LHC reach for stops will push well above the TeV scale, potentially as high as 1.4 TeV. In this case the persistence of null results would generically imply that the weak scale is tuned to the percent level or worse. More optimistically, if SUSY provides a natural explanation for the weak scale, we should be in a position to know during the lifetime of the LHC.

In any event, we are faced with null results in SUSY searches. What are the implications? While we could abandon entirely the prospect for BSM physics near the weak scale, this is needlessly pessimistic. Combining null results with the still-strong motivation of the hierarchy problem instead points us into a variety of interesting directions, many of which entail new (or under-explored) experimental signatures. 
\subsection{Natural SUSY}
An obvious possibility is that our expectations for weak-scale supersymmetry are largely correct, but that the relevant degrees of freedom have evaded us thus far. In particular, it's clear in Fig.~\ref{fig:stops} that there are regions of parameter space where the bounds on stop masses are relatively weak. Populating these regions of parameter space may tell us something about the underlying theory. Broadly speaking, the possibilities are:

\noindent {\it Compressed SUSY:} \cite{LeCompte:2011cn} Limits are weakened if there is a relatively small splitting between the LSP and the top partner; this corresponds to region II in Fig.~\ref{fig:stops}. The small splitting means that hadronic activity in the event decreases. Although the LSPs are each carrying a substantial amount of (missing) energy, the missing transverse energy is reduced and events are difficult to distinguish from SM backgrounds. Sensitivity to these scenarios can be regained in events with hard initial state radiation, where the SUSY process can recoil against more visible energy. 

\noindent {\it Stealth SUSY:} \cite{Fan:2011yu} Alternately, it is possible that SUSY decays can terminate in approximately supersymmetric multiplets, so that there is little true missing energy in the event. In the stop-top system this corresponds to region III in Fig.~\ref{fig:stops}, though more generally there could be additional nearly-supersymmetric multiplets appearing in cascade decays. Stealth decays are extremely difficult to distinguish from Standard Model backgrounds, but precision measurements of the $t \bar t$ cross section and $t \bar t$ spin correlations may provide a handle. 

\noindent {\it Diverse decays:} Most of the sensitivity in Fig.~\ref{fig:stops} is to processes where stops decay into a top plus the LSP. While this decay often dominates, in certain kinematic regimes (labeled region IV in Fig.~\ref{fig:stops}) the preferred decay modes involve an off-shell top. Sensitivity to these decays is somewhat weaker, raising the possibility that stops could be hiding here and motivating searches for these less-distinctive final states. However, in this case the generic relation between the stop and gluino masses requires accommodation of the much stronger gluino mass limits.

\noindent {\it Sflavor structure:} \cite{Blanke:2013zxo} Much of our discussion has assumed that superpartners are approximate flavor eigenstates, so that the decay of the stop proceeds into on- or off-shell top quarks. However, it is also possible for nontrivial mixing between the stop and the superpartner of the charm quark, in which case decays can proceed into a charm quark as illustrated in region IV of Fig.~\ref{fig:stops}. This motivates effective charm-tagging at the LHC. 

\noindent {\it $R$-parity violation:} Finally, all of our discussion thus far has presupposed that stop decays terminate in an LSP stabilized by $R$-parity, giving rise to missing energy signatures. It may instead be the case that $R$-parity is violated, so that decays proceed entirely into Standard Model states. While RPV scenarios are generally no harder to constrain than $R$-parity conserving scenarios, if stop decays proceed predominantly through baryonic RPV operators then they remain essentially unconstrained at the LHC and motivate developing more powerful tools for probing all-hadronic final states.\cite{Bai:2013xla}

\subsection{Unnatural SUSY}

Although it is possible to reconcile natural SUSY with existing limits, the persistence of null results raises a suggestive question: What if SUSY is not completely natural? Supersymmetric extensions of the Standard Model could be moderately tuned while still preserving the other successes of SUSY such as dark matter and gauge coupling unification. In general, the innumerable model-building challenges of weak-scale supersymmetry can be entirely avoided if the mass scale of superpartners is one or more orders of magnitude above the weak scale (albeit at the cost of tuning the weak scale).

Surprisingly, it's not a matter of ``anything goes'' once naturalness considerations are abandoned. In minimal supersymmetric extensions of the Standard Model, the physical Higgs mass leads to a bound on superpartner masses ranging from one to $\sim$ ten orders of magnitude above the weak scale. Moreover, if the success of precision gauge coupling unification is to be preserved, the mass of gauge fermions should not lie more than one or two orders of magnitude above the weak scale. This paradigm of ``mini-split supersymmetry''\cite{Arvanitaki:2012ps}$^,$\! \cite{ArkaniHamed:2012gw} retains a wide variety of novel signatures, particularly involving the (possibly displaced) decays of gluinos and higgsinos.

\subsection{Global symmetries}

Thus far we have focused on SUSY as a solution to the hierarchy problem, but perhaps null results in SUSY searches suggest that a different custodial symmetry is at play in stabilizing the weak scale. The natural alternative is global symmetries (though conformal symmetry is an intriguing and under-explored possibility as well). In the case of global symmetries, the Higgs is identified as a pseudo-goldstone boson of a spontaneously broken global symmetry. The scale of spontaneous global symmetry breaking   must itself be stabilized, either through compositeness or supersymmetry. 

In general, the observation of a light Standard Model-like Higgs implies that such theories should have light fermionic top partners, where the mass scale of the top partners is set by the same naturalness considerations that motivate the mass range of stops. Searches for fermionic top partners generally do not involve missing energy, but rather employ a variety of distinctive Standard Model final states including $tZ, bW,$ and $th$. Current bounds on these states are likewise in the ballpark of 700 GeV, but in general there remain many opportunities for expanding the search for fermionic top partners at the LHC. 

\subsection{Radical symmetries}

\begin{figure}[t] 
   \centering
   \includegraphics[width=2.6in]{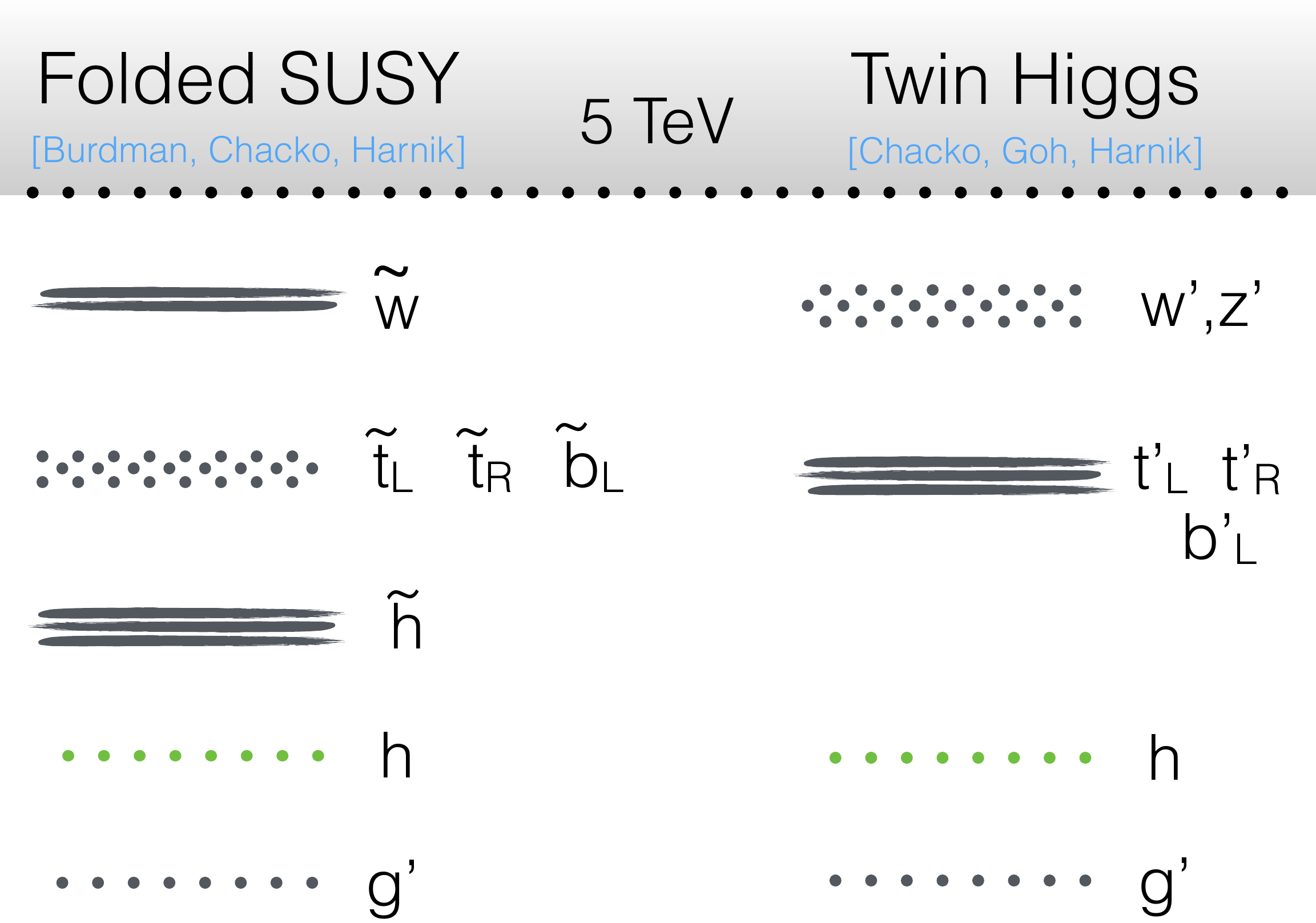} \hspace{5mm}
     \includegraphics[width=3.4in]{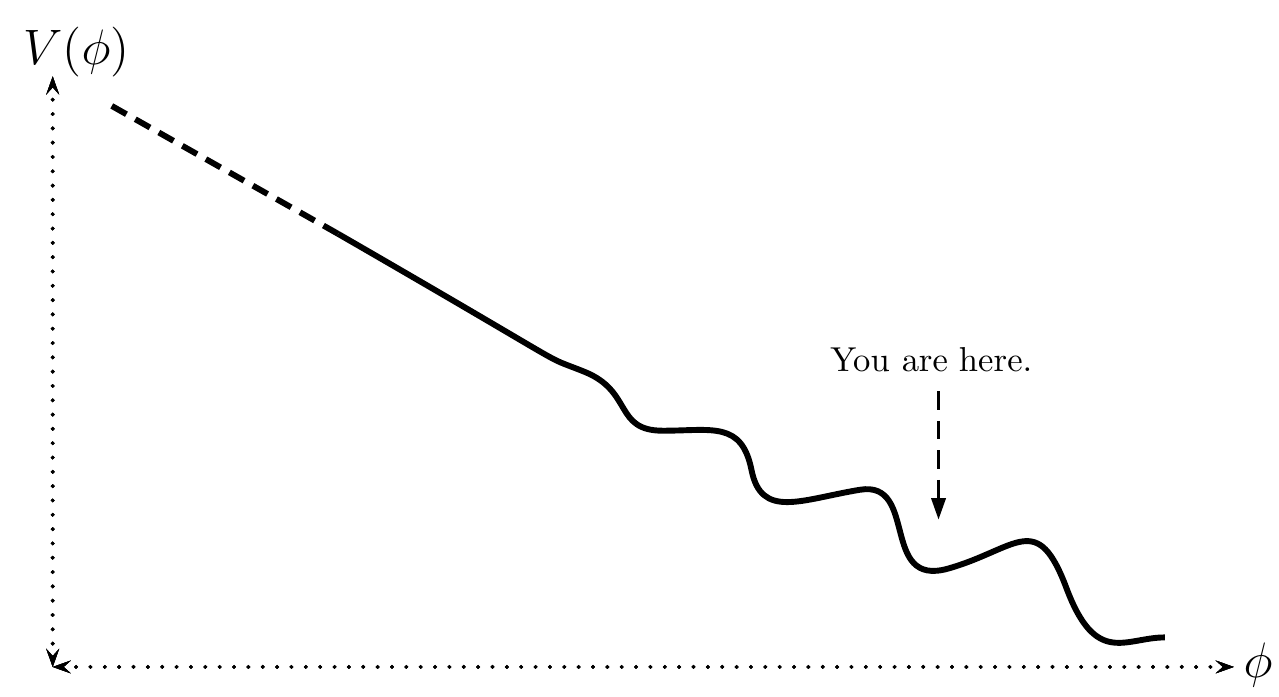}
   \caption{Left: Sample spectra of Folded SUSY and the Twin Higgs. In each case the lightest partner states are neutral under QCD, and in the case of the Twin Higgs are entirely neutral under the Standard Model. Right: Cartoon of the potential for the relaxion. An otherwise smooth potential develops wiggles once $m_H^2 < 0$; dynamical selection of these minima may result in a hierarchy between the weak scale and UV physics in the absence of new weak-scale symmetries.  }
   \label{fig:neutral}
\end{figure}

Both supersymmetry and global symmetries are most tightly constrained by their prediction of QCD-charged top partners, which must be light based on naturalness considerations and (thanks to their QCD charge) should be copiously produced at hadron colliders. However, the production and decay modes used to constrain these top partners are largely separate from the requirements of naturalness, related only by specific custodial symmetries that commute with the Standard Model gauge group. This raises the question of whether it might be possible to have symmetry-based solutions to the hierarchy problem with partner particles that do not carry QCD quantum numbers.

This possibility was first demonstrated in the context of global symmetries in the form of the mirror Twin Higgs \cite{Chacko:2005pe}, while a similar possibility was realized for scalar top partners in the form of Folded SUSY. \cite{Burdman:2006tz} In both cases, the essential symmetry is a discrete symmetry, rather than a continuous one. Such discrete symmetries lead to accidental continuous symmetries for the Higgs mass parameter, where the quantum numbers of the relevant partner states are dictated only by the discrete symmetry and can differ from their Standard Model counterparts. In the case of the Twin Higgs this leads to partner states (such as fermionic top partners) with no Standard Model quantum numbers, while in the case of Folded SUSY it leads to scalar partner particles carrying only electroweak quantum numbers. 

In such theories the discrete symmetry protecting the Higgs relates QCD to a mirror copy of QCD with comparable gauge coupling. This generically leads to exotic signatures involving glueballs of mirror QCD, which can undergo displaced decays at the LHC.\cite{Craig:2015pha}$^,$\! \cite{Curtin:2015fna}
 
\subsection{Not symmetries?}

A final possibility is that the weak scale may be rendered natural without any apparent custodial symmetries. Rather, the weak scale may be selected by some sort of dynamical mechanism. One such way to achieve this is to imagine coupling the Higgs to some field $\phi$ whose minimum sets $m_H^2 = 0$. The challenge is to understand why $m_H^2 = 0$ is a special point of $V(\phi)$. A novel answer to this question was recently provided \cite{Graham:2015cka} where $\phi$ was imagined to be a non-compact axion-like field, the ``relaxion'', coupling to the Higgs. As $\phi$ evolves along its potential, it scans $m_H^2$; when $m_H^2 < 0$, the Higgs acquires a nonzero vev that generates quark masses. These masses in turn generate wiggles in the relaxion potential starting around $m_H^2 \approx 0$. In order for the relaxion to stop in these wiggles, there must be a source of dissipation, which can be provided by inflation. The inflationary sector must be fairly special, with a low Hubble scale and innumerable e-foldings of inflation. The peculiarities of the inflationary sector may merely represent a transferral of tuning in the Higgs potential to tuning in the inflationary potential, but at present this remains an open question.

While in principle such a mechanism might render the weak scale natural without any experimentally-accessible signatures, viable models that significantly raise the cutoff typically involve new ingredients (such as another strongly-coupled gauge group connected to the Standard Model via bi-fundamental matter fields). This in turn implies new naturalness-related physics near the weak scale, though in a form quite different from that encountered in supersymmetry.

\section{Conclusions}

With the discovery of an apparently-elementary Higgs boson at the LHC, the hierarchy problem remains as pressing as ever. Of all the possible solutions to the hierarchy problem, supersymmetry remains perhaps the most strongly motivated explanation but must confront null results from the first run of the LHC. These null results are moderately constraining for supersymmetric scenarios that naturally explain the value of the weak scale. The second run of the LHC (as well as subsequent runs) should prove far more decisive, as the LHC eventually probes superpartners relevant for naturalness above the TeV scale. Whether or not natural supersymmetry lies around the corner, null results should provoke us to think more broadly about possible extensions of the Standard Model. While minimal supersymmetry remains compelling, exploring other solutions to the hierarchy problem opens the door to new searches and signatures at the LHC and other experiments. In the context of natural supersymmetry, null results favor a variety of possibilities ranging from compressed spectra, to approximately supersymmetric (``stealth'') multiplets, to nonstandard sparticle decays or flavor structure, to $R$-parity violation. Looking beyond supersymmetry, null results motivate the exploration of a range of theories involving either alternate custodial symmetries or alternatives to custodial symmetries. We are only beginning to realize the range of possibilities, with much to look forward to in the years to come. 

\section*{Acknowledgments}

This work was supported in part by the Department of Energy under the grant DE-SC0014129.

\section*{References}

\bibliography{blois}

\end{document}